\documentclass[prd, superscriptaddress, tightenlines, longbibliography, nofootinbib, eqsecnum, amsfonts, amsmath, floatfix, notitlepage, twocolumn]{revtex4-2}
\pdfoutput=1

\usepackage[utf8]{inputenc}
\usepackage{mathrsfs}
\usepackage{euscript}
\usepackage{graphics}
\usepackage{graphicx}
\usepackage{amsmath}
\usepackage{amssymb}
\usepackage{gensymb}
\usepackage{bm}
\usepackage[usenames,dvipsnames,svgnames,table]{xcolor}
\definecolor{linkcolor}{rgb}{0.6,0,0}
\definecolor{citecolor}{rgb}{0,0,0.75}
\definecolor{urlcolor}{rgb}{0.12,0.46,0.7}
\usepackage{xspace}
\usepackage{wasysym}
\usepackage{times}
\usepackage{appendix}
\usepackage{comment}
\usepackage{lipsum}
\usepackage[nolist,nohyperlinks]{acronym}
\usepackage{float}
\usepackage{simplewick}
\usepackage{natbib, ifthen}
\usepackage[breaklinks, colorlinks, urlcolor=urlcolor, linkcolor=linkcolor,citecolor=citecolor,pdfencoding=auto]{hyperref}
\usepackage{longtable}
\usepackage{listings}

\defcitealias{PL2018}{PL2018}

\defcitealias{Carron:2022eyg}{PL4}

\newcommand{\isdraft}[1]{}

\newcommand{\av}[1]{\left \langle #1\right\rangle}

\newcommand{\lmax}[0]{\ensuremath{\ell_{\text{max}}}}
\newcommand{\Lmax}[0]{\ensuremath{L_{\text{max}}}}

\begin{document}
\title{Fast partial-sky spherical harmonic transforms}

\newcommand{\vnabla}[0]{\boldsymbol{\nabla}}
\newcommand{\Geneve}{Universit\'e de Gen\`eve, D\'epartement de Physique Th\'eorique, 24 Quai Ansermet, CH-1211 Gen\`eve 4, Switzerland}
\newcommand{\Sussex}{Department of Physics \& Astronomy, University of Sussex, Brighton BN1 9QH, UK}
\newcommand{\MPA}{Max-Planck Institut für Astrophysik, Karl-Schwarzschild-Str. 1, 85748 Garching, Germany}

\newcommand{\Dazro}[0]{\mathcal D_{\valpha=0}}

\author{Julien Carron}
\email{julien.carron@unige.ch}
\affiliation{\Geneve}
\affiliation{\Sussex}
\author{Martin Reinecke}
\email{martin.reinecke@mpa-garching.mpg.de}
\affiliation{\MPA}

  \begin{abstract}
We discuss in some details a novel algorithm for performing partial‑sky spherical harmonic transforms (SHT), building on the Fourier‑sphere method of Reinecke et al.\ (2023) handling efficiently high numbers of arbitrary locations on the sphere.
Our main motivations are Cosmic Microwave Background lensing from the South Pole Telescope, and the South Pole Observatory program targeting primordial gravitational waves from inflation, requiring high-resolution, numerically intensive  work on small sky fractions. 
 We achieve speed-up factors ranging from 3 to 10 on SPT-3G main field and BICEP3 deep footprint, and much more on smaller patches.
More generally, the algorithm eliminates in our case study the usual disadvantages of arbitrary pixelisations in comparison to isolatitude pixelisations or flat‑sky approximations, making it ideal for ambitious workflows that require repeated SHTs on limited sky regions.
  \end{abstract}


   \maketitle
   \tableofcontents
   
 \newcommand{\M}[0]{\textsf{M}}
\section{Introduction}
Let $\theta_p, \phi_p$ be a set of spherical coordinates. We use $\theta$ for co-latitudes and $\phi$ for longitudes. In this paper, we are interested in the problem of synthesizing efficiently and accurately the spin-weighted spherical map 
\begin{equation}\label{eq:syn}
\begin{split}
&f_p=	-\sum_{\ell =0}^{ \lmax} \sum_{m=-\ell}^\ell \left(g_{\ell m}+i c_{\ell m}\right)\:{}_{s}Y_{\ell m}(\theta_p, \phi_p) \\
\\ &\text{(\emph{Synthesis general})}
\end{split}
\end{equation}
at all these coordinates. Here, provided are a set of harmonic coefficients $g_{\ell m}, c_{\ell m}$, obeying the reality conditions $a^*_{\ell m} = (-1)^ma_{\ell, -m}$, up to band-limit $\lmax$, and the spin $s$ is zero or a positive integer. 
On the other hand, provided with an arbitrary set of values $f_p$ (which are complex values for non-zero spins),  we also want to perform the linear operation which is adjoint to \eqref{eq:syn},
\begin{equation}
	\begin{split}
	&\tilde g_{\ell m}=-\frac 12\sum_{p} f_p \:{}_{s}Y^*_{\ell m}(\theta_p, \phi_p)+(-1)^sf^*_p \:{}_{-s}Y^*_{\ell m}(\theta_p, \phi_p) \\
	&\tilde c_{\ell m}=-\frac{1}{2i}\sum_{p} f_p \:{}_{s}Y^*_{\ell m}(\theta_p, \phi_p)-(-1)^sf^*_p \:{}_{-s}Y^*_{\ell m}(\theta_p, \phi_p)
	\\ &\text{(\emph{Adjoint synthesis general})}
	\end{split}
\end{equation}
up to $\lmax$, again efficiently and accurately. These two operations are of fundamental importance in the analysis of data distributed on the sphere, where they can form numerical bottlenecks.
 
 In Ref.~\cite{Reinecke:2023gtp}, we developed a first efficient and accuracy-controlled method in the most general case, when no assumption at all is made on the distribution of the spherical coordinates. However, this method is wasteful when these points are all located within the same portion of the sky. This is the topic of this paper.
 
\begin{figure}
	\includegraphics[width=0.5\textwidth]{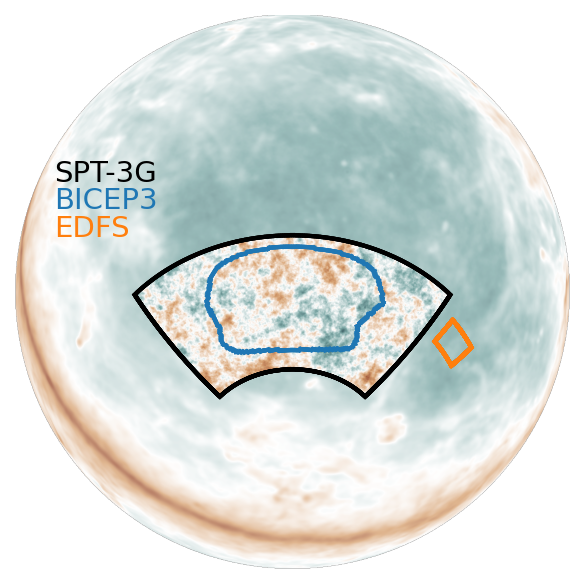}
	\caption{Sky regions we consider here for our tests and that motivated this work.  Black is SPT-3G main field~\cite{SPT-3G:2025bzu}, blue the BICEP3 polarization weight map contour enclosing the central 630 square degrees, and orange the South Pole Telescope observation of the \emph{Euclid} Deep Field South (EDFS)~\cite{SPT-3G:2025rxd}. Their area is about 1500, 800 and 57 square degrees, respectively. \label{fig:dfs_ortho}}
\end{figure}

 Our main motivation was the analysis of deep polarization Cosmic Microwave Background (CMB) data from the South Pole Telescope~\cite{Carlstrom:2009um}. For best results, CMB lensing reconstruction from this data must be performed using iterative methods requiring high numbers of these transforms, on and from set of coordinates without any symmetries. Efficient successful implementation of these methods is also mandatory for the South Pole observational program towards constraining or detecting primordial gravitational waves from inflation from CMB B-modes. See Fig.~\ref{fig:dfs_ortho} for sky regions that motivated this work. The largest of these regions covers about 4\% of the sky. 
 
We discuss in Sec.~\ref{sec:FourierSphere} Fourier series for spherical maps. In~\ref{sec:dfs} we review how they can be computed efficiently for full-sky maps. We then present our new method in Sec.~\ref{sec:small} and show some results. We then conclude in~\ref{sec:conclusions}.
\subsection{Fourier series for spherical maps}\label{sec:FourierSphere}
A spherical map parametrized by $\theta$ and $\phi$ is naturally $2\pi$-periodic in the $\phi$ coordinate. The map can also be easily extended to a periodic map in the $\theta$ coordinate as well~\cite{Risbo1996FourierTransform, Basak:2008pq, Huffenberger:2010hh}. This latter map can be constructed by following great circles, starting, say, from the north pole following a meridian. At the south pole, we simply continue and come back to the north pole from the other side. The angle $\theta$ with its range extended in this way to arbitrary values becomes the natural parameter of the great circle. This operation corresponds to the identification of the standard spherical coordinate $$(\theta=\pi-\beta, \phi + \pi),$$ reached after crossing the south pole, to the $\theta$-extended coordinate $$(\theta=\pi + \beta, \phi).$$

There is one slight twist to consider in the case of spin-weighted fields. Very close to the pole, as one goes around the pole, spin-weighted fields are proportional to $e^{i s\phi}$, owing to the rotation of the local basis axes $\boldsymbol{e}_\theta$ and $\boldsymbol{e}_\phi$. Hence, for continuity, we must include a sign $(-1)^s$ when crossing the pole.

One can also see this more formally. In the appendix, we argue that spherical harmonics can be understood without restrictions to $\theta$ and $\phi$, with the rule

\begin{equation}\label{eq:sht}
	{}_{s}Y_{\ell m}(\pi + \beta, \phi) =  (-1)^s{}_{s}Y_{\ell m}(\pi-\beta, \phi+\pi),
\end{equation}
As a consequence, the spherical map
\begin{equation}\label{eq:f}
	{}_{s}f(\theta, \phi) = -\sum_{\ell=0}^{\lmax}\sum_{m=-\ell}^\ell \left(g_{\ell m} +ic_{\ell m}\right) {}_{s}Y_{\ell m}(\theta, \phi)
\end{equation}
is well-defined for any values of $\theta$ and $\phi$, and is periodic in both coordinates, with period $2\pi$. 
\newline
Hence, spherical maps can be tentatively assigned a 2D Fourier series. It is less obvious, but a crucial  point, that the Fourier series is also band-limited when the spherical expansion is, and with the exact same band-limit. It follows that spherical maps band-limited to $\lmax$ also have an exact representation
\begin{equation}	
{}_{s}f(\theta, \phi) = \sum_{n=-\lmax}^{\lmax}\sum_{m=-\lmax}^{\lmax}  \tilde f_{n m}\: e^{i n \theta + i m \phi}
\end{equation}
for some set of Fourier coefficients. See the appendix.
\subsection{Evaluation at any point from iso-latitude rings}\label{sec:dfs}
Following the discussion above, the problem of evaluating a spherical map at arbitrary locations may be turned into that of evaluating a standard, band-limited Fourier series on these locations.
Hence, for a band-limit $\lmax$, sampling theorems ensure that the values of $f$ at $2\lmax + 1$ $\phi$-points $\times$ $2\lmax + 1$ $\theta$-points contain all the necessary information. Furthermore, the symmetry of the doubled Fourier sphere\footnote{The equivalence of $(\pi - \beta, \phi + \pi)$ to $(\pi + \beta, \phi)$, up to a sign.} halves the number of independent $\theta$ points. Hence, in order to be able to evaluate $f$ exactly at any point, knowledge of $f$ at only $\lmax + 1$ co-latitude rings with each $2\lmax + 1$ points is necessary.

Say we have $2\lmax + 1$ values at equidistant points $\theta_j$ of a one-dimensional band-limited function with band-limit $\lmax$. For example, values of the spherical map along a meridian. The desired exact value of the function at $\theta$ is then given by computing the non-zero frequencies from the points, and Fourier-sum them back to $\theta$. This corresponds to the application of the kernel
\begin{equation}\label{eq:Dnz}
D_j^\theta \equiv \frac{1}{2\lmax + 1}	\sum_{m=-\lmax}^{\lmax} e^{im(\theta-\theta_j)}
\end{equation}
to the set of points: it holds namely
\begin{equation}
\sum_{j=-\lmax}^{\lmax} D_j^\theta f(\theta_j) = f(\theta).
\end{equation} 
In practice, this can be computed with an FFT followed by a non-uniform FFT, for which very efficient algorithms exist, including~\cite{Barnett} or \texttt{ducc0.nufft}\cite{2020ascl.soft08023R}. In general, 2D FFT followed by 2D non-uniform FFT provides efficient evaluation of a spherical map from the full set of $\lmax + 1$ rings. This is the essence of the method described in~\cite{Reinecke:2023gtp}. Non-uniform FFTs incurs errors beyond those sourced by floating-point arithmetics. These errors are controlled by a user-specified tolerance~\cite{Arras, BarnettESaliasing}.

\section{Spherical interpolation on small patches}\label{sec:small}

\begin{figure}[h!]
	\includegraphics[width=0.49\textwidth]{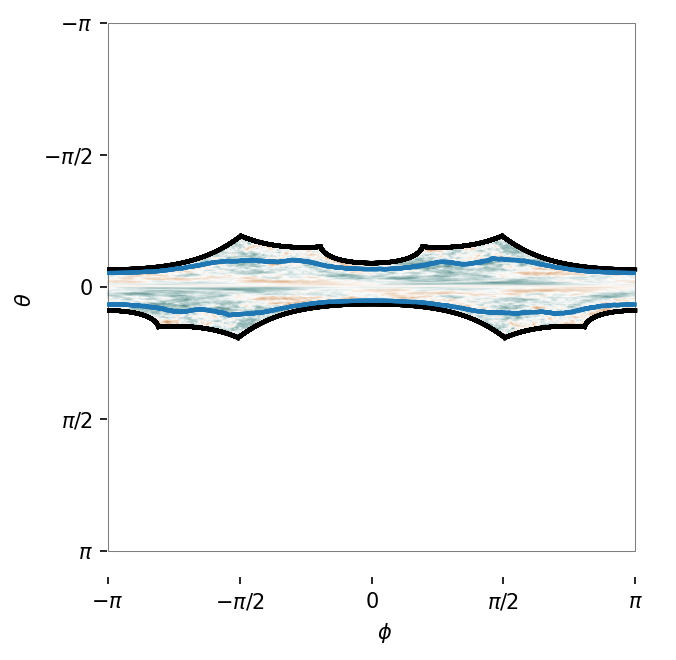}
	\includegraphics[width=0.49\textwidth]{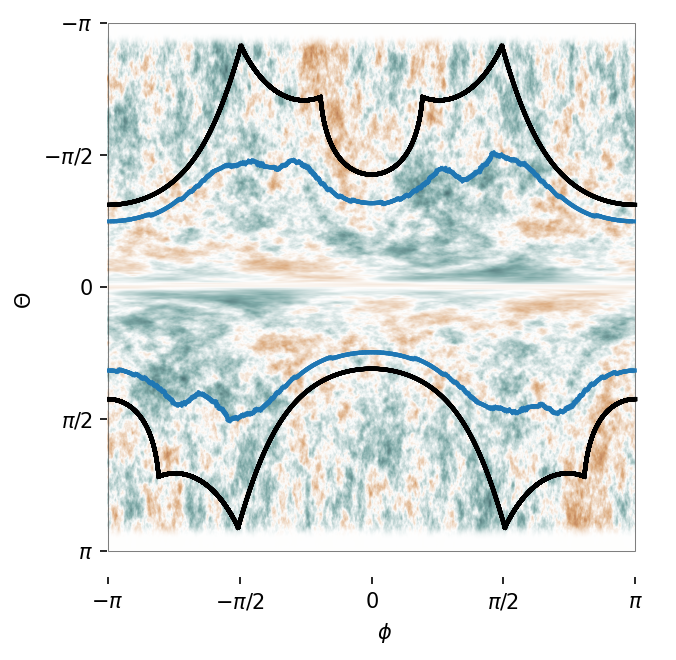}
	\caption{Illustration of the \emph{synthesis general} operation, for a scalar transform. The top panel shows the map to be evaluated on a set of points of sphere, starting from given spherical harmonic coefficients. In this case, the region of interest is the SPT-3G region as defined by the black contours on Fig.~\ref{fig:dfs_ortho}. This is shown after placing the north pole $\theta=0$ in the center of the patch. Placing the pole in the center stretches the map to a maximum horizontally, reducing the band-limit in that direction (reducing $m_{\text{max}}$). The lower panel shows the Fourier map which is actually evaluated, after stretching and weighting in the vertical direction. The effective Fourier band-limit of the map is now reduced to a minimum in both directions, allowing faster evaluation. In the case of BICEP3, in blue, we can visibly stretch significantly more, improving performance further.\label{fig:steps}}
\end{figure}
Let us first recap the previous section: we can evaluate a spherical map at large numbers of arbitrary points by combining efficient Fourier techniques (uniform and non-uniform Fourier transforms), with the direct evaluation of the map on $\lmax + 1$ equidistant rings only. Since the numerical cost of evaluating a spherical map is vastly dominated by the Legendre transform, and hence by the number of different $\theta$ values, this approach is plausibly close to optimal in many practical cases -- it is not possible to use fewer co-latitude values than $\lmax + 1$ for an exact evaluation of $f$.

However, consider now that all coordinates of interest are within a small region of the sky. If the region is really small, most rings will be far away from it, and clearly the method is wasteful.

Instead, if the small region could be extended slightly to a new effectively periodic function (of smaller period), then we could reuse the same machinery, but computing the function at $\theta$ values only inside the region. We would need then a much smaller number of $\theta$ points than $\lmax + 1$ in order to probe the same physical features.

Equivalently, we can imagine stretching the small area, making it bigger such that it occupies a larger part of the Fourier sphere. This has the effect of magnifying all features, reducing effectively the band-limit of the map by making them larger. Hence we also need less rings than before in this picture. This is the picture we adopt.
\\ 
Of course stretching the patch affects the Fourier content of the map in a non-trivial manner, and we must make sure the resulting map is periodic and as smooth as possible. Hence we cannot expect to provide exact results anymore. However, as we show, error can be monitored simply and can be made as small as desired.
\\
\newcommand{\thtmax}[0]{\ensuremath{\theta^{\textrm{max}}}}
\newcommand{\epsa}[0]{\ensuremath{\epsilon^{\textrm{apo}}}}
We proceed as follows, explaining some aspects later in more detail:
\begin{enumerate}
	\item We use coordinates such that the center of the patch is at the north pole. In these coordinates, we now define $\theta^*$ such that all points of interest are in the spherical cap defined by $|\theta| \le \theta^*$. 
	\item We then choose an apodization length, and make the spherical cap slightly larger, now extending to
	\begin{equation}
		\theta^{\rm max} \equiv \theta^*(1+\epsa)
	\end{equation} 
	\item We pick a weight function $w$ and define a weighted, $\theta$-remapped function
	\begin{equation}
		\tilde f(\Theta, \phi) 
	\equiv  w(\Theta)f\left(\theta=\Theta \frac \thtmax \pi, \phi \right)
	\end{equation}
	To avoid ambiguity, we use $\theta$ for coordinates in the original small patch, and $\Theta$ after enlarging the patch to the full Fourier sphere.
		The weight $w$ is chosen to be exactly equal to unity for $\theta \le  \theta^*$, and decaying smoothly until it vanishes at $\thtmax$. 
	\item The full-sky method of Ref.~\cite{Reinecke:2023gtp} is then used on this new function, using a band-limit $\Lmax$: we obtain $f(\theta, \phi)$ by evaluating $\tilde f(\Theta, \phi)$. This requires evaluation of $\Lmax + 1$ iso-latitude rings of the original function $f$. All of these rings are within the spherical cap defined by $\thtmax$.
\end{enumerate}
See Fig.~\ref{fig:steps} for an illustration. The size of the apodization region $\epsa$, the weight function $w$, and the effective new band-limit $\Lmax$, are parameters that must be chosen jointly. The aim is to make the computational cost as small possible, with the constraint that the error on the result must be within some tolerance $\epsilon$.

\subsection{Expected performance}
\begin{figure}
	\includegraphics[width=0.5\textwidth]{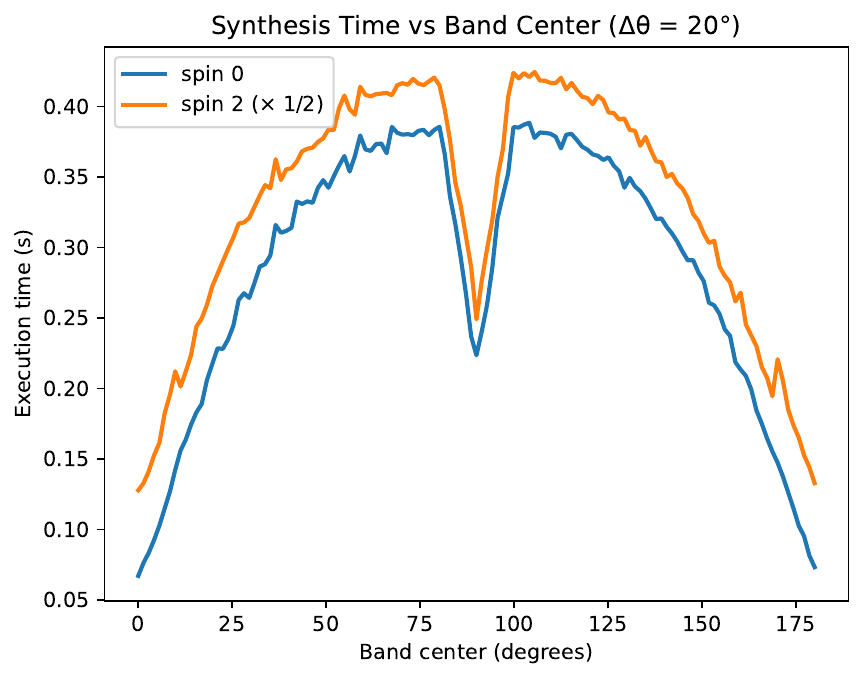}
	\caption{\emph{Polar optimization and equatorial symmetry: }Performance comparison of \texttt{ducc0} spherical harmonic synthesis on regions of fixed co-latitude extent (here 20 degrees), as a function of the center co-latitude. Each region has the same number of pixels, which is set very small for this test. Shown are spin-0 and spin-2 transforms (the latter rescaled by a factor $1/2$) , using $\lmax = 5000$. Owing to the irrelevance of $m$-values well above $\lmax \sin \theta$, optimal performance is reached near the poles. The trough at the equator arises when the band crosses the equator: the underlying $\theta$ grid is such that it is symmetric with respect to the equator, and the calculation of a pair of symmetric rings has the same cost than for one member of the pair. \label{fig:polar}}
\end{figure}
We now discuss crudely the expected performance gain of the method. The dominant source of numerical cost in these operations is that of the calculation of the Legendre (or Wigner) coefficients for each $\theta$. In our approach, there are two conceptually distinct contributions towards alleviating this cost:
\begin{itemize}
	\item There are less iso-latitude rings to compute ($\Lmax + 1$ versus $\lmax + 1$) 
	\item  Each ring is within a spherical cap, and some will be very close to the pole. Rings close to the pole are faster to compute. This is because the physical length of the ring is reduced by $\sin(\theta)$ compared to the equator. This makes longitudinal waves $e^{im \phi}$ of frequency $m$ much larger than $\lmax\sin(\theta)$ irrelevant -- large $m$ values can be dropped altogether (reducing the range of $m$ computed as function of $\theta$ in this way is called \emph{polar optimization}~\cite{libsharp}, see also Fig.~\ref{fig:polar}).
\end{itemize}

Hence, the Legendre cost associated to the $\Lmax + 1$ rings may be written
\begin{equation}
	\text{Legendre cost} \propto 2\pi \sum_{j=0}^{\Lmax} \sin\left(\theta_j=\Theta_j \frac{\thtmax}{\pi}\right),
\end{equation}
with
\begin{equation}
	\Theta_j = j\frac{\pi}{\Lmax}, \quad j = 0, \cdots, \Lmax
\end{equation}
If we interpret this sum as an integral, we get
\begin{equation}
	\text{Legendre cost} \propto \left(\frac{\Lmax}{\thtmax} \right) \text{Cap Area}(\thtmax),
\end{equation}
The Legendre cost of the full-sky method is obtained by using $\lmax$ in place of $\Lmax$, no rescaling, and integrating to $\pi$ instead of \thtmax.
There is one difference, which is that owing to the symmetry of the rings with respect to the equator, the number of independent rings to compute is halved (see the trough in the center Fig.~\ref{fig:polar}). Thus we expect the Legendre part to be faster by about half the inverse sky fraction of the spherical cap,

\begin{equation}
\begin{split}
	\left(\text{Legendre speed-up}\right)^{-1} &\approx\left(\frac{\pi}{\thtmax}\frac{\Lmax }{\lmax}\right)2f_{\text{sky}}^{\text{cap}}(\thtmax)\\
\end{split}
\end{equation}
The prefactor in brackets must be unity plus a correction (it will be for us in fact exactly equal $1 + \epsa$ later on). The improvement is carried by the reduction in area, from the full north hemisphere to the spherical cap. The area of the spherical cap defined by $\thtmax$ can be significantly larger than the area of the patch itself -- the two areas are only very close to each other for round regions.\\
The perfect polar optimization scaling assumed here is optimistic, and in practice tiny patches do not benefit from it quite as much. In the absence of any polar optimization, the expected gain is obtained by replacing the sky area of the spherical cap by the area of the rectangular band defined by $\thtmax$ on the Fourier sphere (with scaling $\propto \thtmax$, instead of $(\thtmax)^2$ for small spherical caps)

\subsection{Choice of apodization length}
We discuss now our choice of apodization length. The argument as presented here is somewhat heuristic but works very well in practice and can be confirmed a posteriori.\\
 There are two competing effects of the apodization length that impact \Lmax. On one hand, a small apodization length is desirable, since we can then stretch more the original patch, magnifying more the features inside the patch of interest. On the other hand, a large apodization length is desirable, to avoid sharp features in the transition region.
\\
Let $\Delta \lmax$ the increase in effective band-limit induced by the weighting. Let $\Delta\theta$ the increase in patch size. We expect roughly
\begin{equation}
\begin{split}\label{eq:Lmax}
	\Lmax &\sim  \left(\lmax + \Delta\lmax\right)\frac{\theta^* + \Delta \theta}{\pi} \\&= \left(\lmax \frac{\theta^*}{\pi}\right)\left(1 + \frac{\Delta\lmax}{\lmax}\right)\left(1 + \frac{\Delta \theta}{\theta^*}\right).
\end{split}
\end{equation}
We can go further relating $\Delta \lmax$ to $\Delta \theta$. Say one needs an effective number of modes $\Delta \ell$ to describe adequately a transition function on an interval of size $\pi$. After rescaling to the apodization interval $\Delta \theta$, the increase to the band-limit will be
\begin{equation}
	\Delta \lmax = \Delta \ell \frac\pi{\Delta \theta}.
\end{equation}

Our choice of weight function will only depend on the desired precision of the result. Hence we can use this relation to try and minimize the numerical cost for a given $\lmax$ and $\thtmax$. A good approximation is simply to minimize $\Lmax$.
Optimizing~\eqref{eq:Lmax} results in
	\begin{equation}   \epsa \equiv \frac{\Delta \theta}{\theta^*} = \sqrt{\frac{\pi\Delta \ell}{\theta^* \lmax}}=\frac{\Delta \lmax}{\lmax}.
\end{equation}
The resulting band-limit is
\begin{equation}\label{eq:Lmax2}
	\Lmax = \left(\lmax \frac  {\theta^*}\pi\right)(1 + \epsa)^2.
\end{equation}
The relative increase in the number of points compared to the best case scenario is $(1 + \epsa)^2$\footnote{ 
Instead of minimizing $\Lmax$ as we did, we can also minimize the expression for the Legendre cost inclusive of polar optimization. This is forecast to gives very small further gains (x$\sim 2\%$ for SPT-3G main field) for the main motivations of this work. A significant difference is expected only in the limit of tiny patches}.
The only remaining parameter, $\Delta \ell$, is set by the window. It stands in one-to-one correspondence to the desired accuracy of the result. We discuss this now.

\subsection{Choice of window}

\begin{figure}[h]
	\includegraphics[width=0.95\columnwidth]{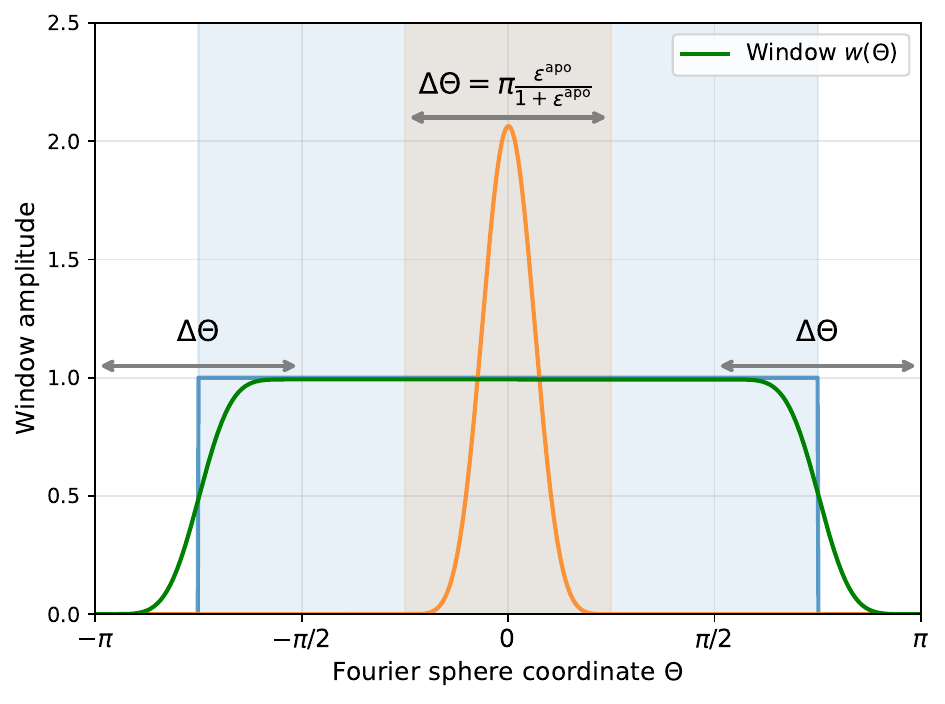}
	\includegraphics[width=0.95\columnwidth]{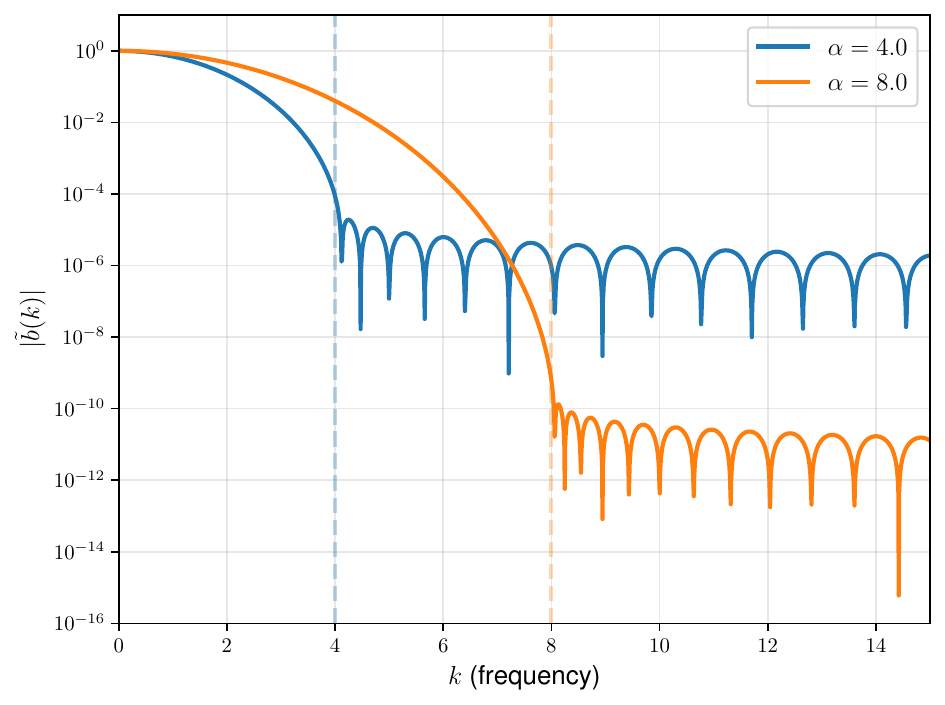}
	\caption{\emph{Upper panel}: Construction of our window, green, from the convolution of a top-hat, in blue, with a compactly supported bell-like function, $b$, orange (here the ES kernel, indistinguishable from a Gaussian on this figure), rescaled such that its support is $\Delta \Theta$. The support of the top-hat is $2\pi-\Theta$.  In this construction, the rise and decay of the window is given by the cumulative function of the bell.
	In Fourier space, the window is in all cases the same universal function at the rescaled frequencies $k = n \Delta\Theta /2\pi$, given by the product of the two Fourier windows. The problem of optimizing the window is reduced to that of picking the best bell function.\\
	\emph{Lower panel:} Fourier transform of the Slepian-alike Kaiser-Bessel bell, for two values of shape parameter (main lobe width) $\alpha$, as given by Eq.~\eqref{eq:KB}. Desirable is $\alpha$ as small as possible, reducing the width of the main-lobe and the corresponding apodization length. However, this is only possible as long as the side-lobe of the window is well below the desired tolerance $\epsilon$. This reasoning gives the direct correspondences for the best choice of window $\pi \alpha \sim |\ln \epsilon|$, and $2\alpha \sim \Delta \ell$.
	\label{fig:convolution} }
\end{figure}

Here we describe the construction of our window functions. We would like a smooth enough function, which is unity on the interval of interest, and then decays to zero on an interval sized $\Delta \Theta$. A simple way to form such a function is by the convolution of a smooth `bell-shaped' function, of support $\Delta \Theta$ and unit area, with a top-hat of support $2\pi-\Delta\Theta$. 
In this construction, the rise and decay of the window at the endpoints are given by the cumulative function of the bell. See Fig.~\ref{fig:convolution}.
\\Further, the window must have minimal frequency content, in order to provide a an effective band-limit to the weighted function as small as possible. Since the window is a convolution in real space, its Fourier transform will be product of that of the bell with that of the top-hat. The latter Fourier transform decays slowly. Hence, we are interested in compactly supported bells with Fourier content decaying as quickly as possible.

By the uncertainty principle, a function cannot be simultaneously compactly supported in both position space and the frequency domain, so exact compact support in one domain forces infinite support in the other. One classical formulation of this problem leads to the study of functions that are strictly compactly supported in one domain while optimizing the concentration of their content in the other within a prescribed range~\cite{Slepian}. This variational problem gives rise to the prolate spheroidal wave functions (also known as Slepian functions), as one possible optimal compromise between these conflicting requirements~\cite{Slepian}. In our case they were found to outperform all other classes of functions we tried, often by a wide margin.
\\
Relevant are Slepian functions of order 0. They have one shape parameter, `$\alpha$', the frequency range on which their power is concentrated. They do not have simple closed forms. For increasing $\alpha$, they converge to Gaussians with variance $1/\alpha$ away from the boundaries. In the Fourier domain they also first form a Gaussian-like main lobe. At frequency $\alpha$, then they start to decay much more slowly.
\\Several convenient and good approximations to these functions are well known, including the Kaiser-Bessel window~\cite{KaiserBessel} which has a useful closed form Fourier transform (see lower panel Fig.~\ref{fig:convolution}), given by
\begin{equation}
\begin{split}
\tilde b(k) &\equiv \int_{-\pi}^{\pi} d\Theta\: b(\Theta)e^{-i k \Theta} \\ &=\frac{\pi \alpha}{\sinh(\pi \alpha)}\frac{\sinh(\pi \sqrt{\alpha^2-k^2})}{\pi\sqrt{\alpha^2 - k^2}}
\end{split}
\end{equation} 
We ended up using the `exponential of semi-circle' bell introduced in \cite{Barnett} in the context of non-uniform Fast Fourier Transforms, where functions with similar requirements are essential,
\begin{equation}\label{eq:KB}
	b(\Theta) \propto \exp\left[\alpha\pi\left(\sqrt{1- (\Theta/\pi)^2}-1\right)\right],
\end{equation}
which gives only minimally worse results, compared to the Kaiser-Bessel or Slepian bells. The window behaves to a good approximation as $\sim e^{-\alpha \pi}$ both at the interval end-points in real space, and at the side-lobe in Fourier space. Reasoning as in the caption of Fig.~\ref{fig:convolution} gives the following rule for the choice of window,
\begin{equation}
	\alpha = - \frac{\ln \epsilon}{\pi}, \quad \Delta\ell=2\alpha,
\end{equation}
which works very well in practice.\\
\textbf{Summary:}
We are now in position to summarize our choice of parameters (up to possible additional small fine-tuning):
Given the input tolerance $\epsilon$,  we pick the bell in the Slepian-like family with main-lobe width 
\begin{equation}
	\alpha =  \frac{|\ln \epsilon|}{\pi}.
\end{equation} 
The fractional apodization length is then
\begin{equation}
		\epsa = \sqrt{\frac{2|\ln \epsilon|}{\theta^* \lmax}}.
\end{equation}
The new band-limit $\Lmax$ follows then from Eq.~\eqref{eq:Lmax2}.
\subsection{Errors}
There are two main elements defining the precision of the result. One is the non-uniform Fast Fourier Transform (nuFFT) implementation. Our work brings nothing new in this respect, and we refer to the nuFFT literature~\cite{Barnett, BarnettESaliasing}.
The other source, new to this work, is that the function which we evaluate, $\tilde f$, is different to the one we are really interested in, $f$. We must make sure the difference is small enough. We discuss this now in more details.\\
Consider a wave along the latitude direction $e^{i m \theta}$ of frequency $m$  on the Fourier sphere. 
After stretching and weighting, this wave appears on the new Fourier sphere as an apodized wave of lower frequency:
\begin{equation}\label{eq:wave}
	\exp\left(im \theta\right)\rightarrow w(\Theta) \exp \left(i M\Theta\right),
\end{equation}
where \begin{equation}
	M \equiv m  \left(\frac{\thtmax}{\pi}\right),\quad \Theta \equiv  \theta \left(\frac {\pi}\thtmax \right).
\end{equation}
The effective band-limit on the right-hand side of~\eqref{eq:wave} is taken to be $\Lmax$.  Error will occur if the band-limited approximation to the low-frequency wave is not accurate.\\
Explicitly, the error made on the wave will be given by
\begin{equation}\label{eq:s2m}
\sigma^2_m(\theta)\equiv \left|\left(\sum_{j=-\Lmax}^{\Lmax}w_jD^\Theta_j e^{i m \theta_j}\right) - e^{im \theta} \right|^2
\end{equation}
where the kernel $D$ is the one given by Eq.~\eqref{eq:Dnz}, which computes the band-limited approximation of the wave from its values at the given points $\theta_j$ on the new Fourier sphere,
\begin{equation}
	D^\Theta_j \equiv \frac{1}{2\Lmax + 1}\sum_{M=-\Lmax}^{\Lmax}e^{i M(\Theta-\Theta_j)}.
\end{equation}
Eq.~\eqref{eq:s2m} can be computed directly efficiently\footnote{For example with non-uniform fast Fourier transforms}. As naively expected, the error is larger for $\theta$ close to the transition region\footnote{By construction, the error is zero when $\Theta$ equals one of the $\Theta_j$, and takes maximal values half-way  between those points. These statements apply to the error envelope.}, and larger at high frequencies.\\
\begin{figure}
	\includegraphics[width=0.98\columnwidth]{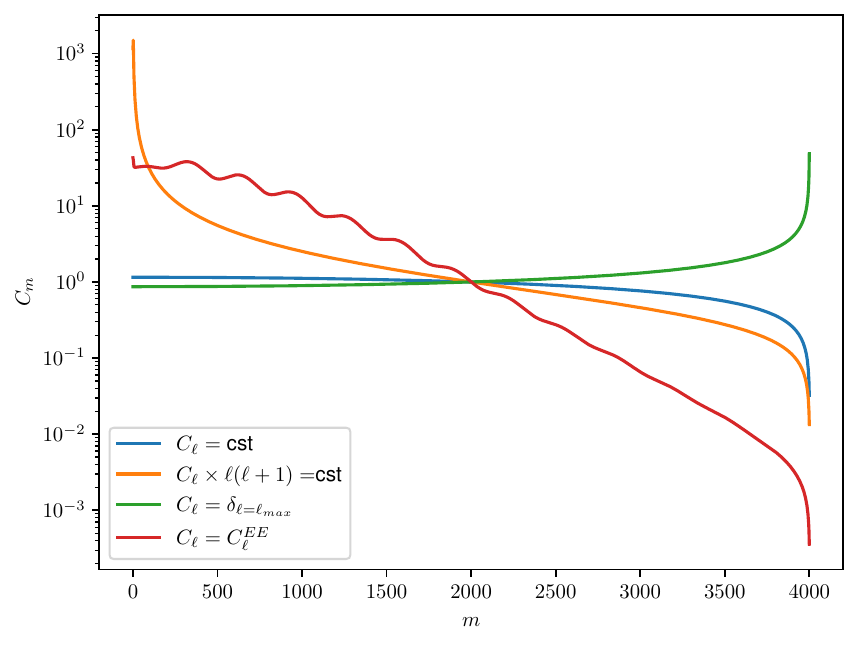}
	\includegraphics[width=0.98\columnwidth]{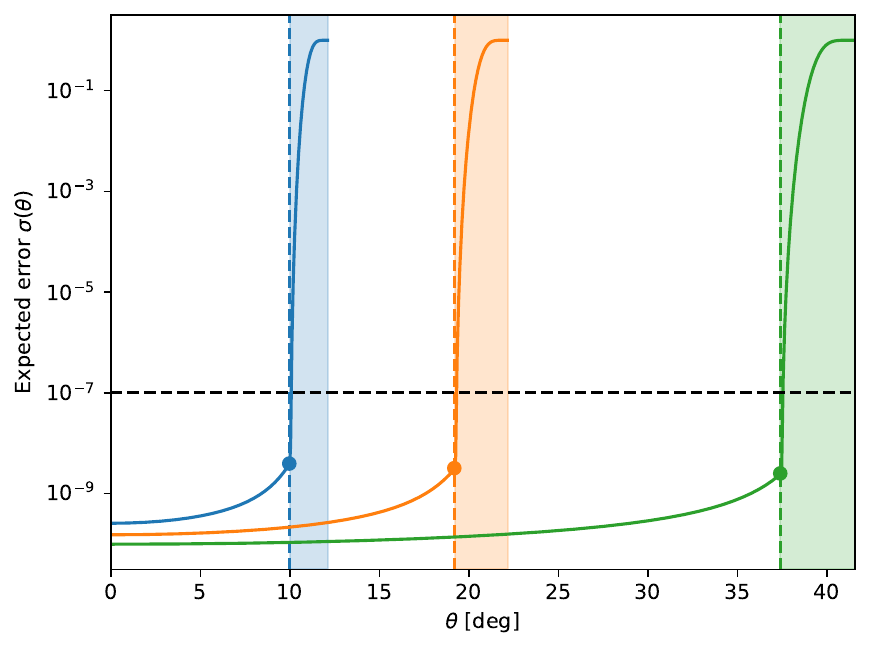}
	\caption{\emph{Top panel:} One-dimensional power spectra $c_m$ along great circles for a few special cases of spherical power spectra $C_\ell$. This is for $\lmax = 4000$, and with arbitrary normalization. Blue has a perfectly white spherical spectrum, and orange a perfectly scale-invariant one. Green shows the case where there is power only at $\lmax$ and nowhere else. Red results from a $\Lambda$CDM polarization $E$-mode power spectrum.
	\emph{Lower Panel:} Predicted (envelope of) the error, Eq.\eqref{eq:st}, as function of $\theta$ in our baseline apodization procedure, for three spherical caps $\theta^*$ as indicated by the vertical dashed lines. The colored bands indicate the apodization region. The circles indicate the worst error inside the cap. This is for $\lmax = 4000$ and desired tolerance $\epsilon=10^{-7}$ (dashed black), and for the case of a perfectly white spherical power spectrum. }
	\label{fig:cms} 
\end{figure}
Let us consider now the case of statistically isotropic fields, which is most relevant for CMB analysis. The map of interest along a great circle of constant $\phi$ is also a (one-dimensional) statistically isotropic field. The two-point correlations along the great circle can be described with the help a one-dimensional power spectrum $c_m$,
\begin{equation}
	\av{f(\theta, \phi)f^*(\theta', \phi)} =\frac1 {2\pi} \sum_{m=-\lmax}^{\lmax} c_m e^{im (\theta-\theta')}
\end{equation}
The error that we are making by assuming a band limit $\Lmax$ will then be given by collecting the errors made on all the waves,
\begin{equation}\label{eq:st}
	\begin{split}
	&\av{\left|\tilde f(\Theta, \phi) - f(\theta, \phi)\right|^2} \equiv \sigma^2(\theta)\\&= \frac{1}{2\pi}\sum_{m=-\lmax}^{\lmax} c_m \sigma^2_m(\theta).
	\end{split}
\end{equation}
Obviously, the error we are making depends on the input field. Since the error on the waves is located mostly at high-$m$, the final error will be disproportionally smaller for smoother fields with little small scale $m$-power. 
In the case of scalar transforms, this 1D power along great circles is related to the spherical power $C_\ell = \av{\left|g_{\ell m}\right|^2}$ by
\begin{equation}\label{eq:cm}
\begin{split}
	c_m &\equiv  \sum_{\ell=m}^{\lmax} C_\ell \frac{2\ell+1}{2} \left|d^\ell_{m0}\left(\frac \pi 2\right)\right|^2, \quad (\text{spin }s=0)\\
	&\approx \frac{1}{\pi} \int_{m}^{\lmax} d\ell\:\frac{C_\ell}{\sqrt{1 - (m/\ell)^2}}
\end{split}
\end{equation}
The first line is exact, the second very accurate for high moments (`flat-sky' approximation).
The case of power being concentrated at the highest $m$ is unnatural for spherical maps: the number of $\ell$-multipoles above $m$ and below $\lmax$ gets smaller, and this will cause $c_m$ to drop in most cases of relevance. See Fig.~\ref{fig:cms}.\\ 

For spin-weighted transforms, the corresponding relation is
\begin{equation}\label{eq:cms}
	c_m \equiv  \sum_{\ell=m}^{\lmax} \left(C^E_\ell + C_\ell^B\right) \frac{2\ell+1}{2} \left|d^\ell_{m-s}\left(\frac \pi 2\right)\right|^2, \quad (s>0),
\end{equation}
where $C^E$ and $C^B$ are the spectra of the gradient and curl mode of the spin-weighted field. The high-moment limit is the same as above. Eq.~\eqref{eq:st} gives the squared magnitude of the error. 
Errors in the real and imaginary part of the field can be different. Let the spin-$s$ field be\begin{equation}
	{}_{s}f(\theta, \phi) = (Q +iU)(\theta, \phi).
\end{equation} In the case of a pure gradient ($E$-mode) field, the high-momentum $c_m$ of $Q$ and $U$ for great circles of constant $\phi$ will be given by
\begin{equation}
\begin{split}
	c_m&\approx \frac{1}{\pi} \int_m^{\lmax} d\ell \frac{C^{E}_\ell}{\sqrt{1 - (m/\ell)^2}} \cos^2(s\varphi) \quad (Q)\\
	c_m &\approx \frac{1}{\pi} \int_m^{\lmax} d\ell \frac{C^{E}_\ell}{\sqrt{1 - (m/\ell)^2}} \sin^2(s\varphi),\quad (U)
\end{split}
\end{equation} where $\varphi$ is defined through $\cos \varphi \equiv m /\ell$. For our typical applications, this implies that errors for $Q$ are noticeably larger than for $U$.\\
In general, $c_m$ can be computed very efficiently for any $C_\ell$ for example from the one-dimensional Fourier transform of the real space correlation function.
\subsection{Some results}
\begin{table*}[htbp]
\centering
\caption{Some performance benchmarks. The number of points is set by the effective resolution in the fourth column (corresponding crudely to that of an Healpix~\cite{Gorski:2004by} nside parameter of 2048, 4096 or 8192). Speed-up factors compared to original \texttt{ducc0.sht 0.40.0} routines are given in the 5th and 6th columns. They  are computed from 10 realizations, each using  4 cores. Our new method is implemented as minor tweaks to the most recent \texttt{ducc} code (only the weighting of the rings and their locations are new). All other elements are identical, benefiting from the same CPU vector accelerations and other optimizations. Here the tolerance is set to $10^{-7}$. The tolerance only has a minor impact on the execution times -- this only affects a fractional correction to the total work to perform. SPT-3G+/BICEP3 denotes a subregion of SPT-3G footprint enclosing but slightly larger than the BICEP3 field.}
\label{tab:syng_cap_benchmark}
\begin{tabular}{|l|r|r|r|r|r|r|rr|}
\hline\hline
Configuration (area) & $\theta^*$ [deg.] &$\ell_{\max}$ & resolution [amin] & Spin  & \emph{Synthesis general} & \emph{Adjoint Synthesis general}& predicted speed-up &\\ 
\hline
SPT-3G Main (1500 $\text{deg}^2$) &$35^\circ$ & 4000   &1.7 & 0 & 3.5 & 3.3 & 4.1 & \\
 && 4000   &1.7 & 2 & 3.2& 3.1 & 4.1 &\\
  && 5120   &1.7 & 1 & 3.2&  3.4&4.2\\

 && 6000   &1.7 & 0 & 3.9&  3.8&4.3\\
 && 10000   &0.43 & 0 & 3.5& 3.3 & 4.6 &\\ 

SPT-3G+/BICEP3 (980 $\text{deg}^2$)  &$20^{\circ}$& 4000  & 1.7 & 0  & 7.7 & 7.12 & 11.5 &\\
  && 4000  & 1.7 & 2  & 6.4 & 6.5 &11.6& \\
  && 6000  & 1.7 & 0  & 9.0& 8.8 & 12.5 &\\

EDFS (57 $\text{deg}^2$) &  $7^{\circ}$ & 4000 & 0.86 & 2  & 23& 21 & 67& \\
& & 10000 &0.43 & 0  & 39& 43 & 85& \\
\hline\hline
\end{tabular}
\end{table*}
These prescriptions require only a small number of tweaks to the currently best-performing SHT code  \texttt{ducc0.sht}\footnote{\url{https://gitlab.mpcdf.mpg.de/mtr/ducc/-/tree/ducc0}}. The only new pieces are the reinterpretation of the coordinates ($\theta\rightarrow \Theta$) and the weighting of the Fourier sphere rings by $w(\Theta)$, which are simple -- everything else stays identical, benefiting from the same carefully aggressive optimization.\\

In Table~\ref{tab:syng_cap_benchmark} we discuss some results, comparing \emph{synthesis general} and  \emph{adjoint synthesis general} to this new version. 
We see speed-up factors of about 3-4 for SPT-3G, 6 to 9 for BICEP3, and much more in the case of much smaller EDFS field (up to 40). The speed-ups are a bit smaller than discussed earlier in this paper, because of our optimistic treatment of polar optimization (and to a smaller extent from the fact that the Legendre part is only a part of the computation). 

Here is a couple of additional comments, showing how this method can remove requirements of both isolatitude pixelizations (or a pixelization at all), and of flat-sky approximations.
\begin{itemize}
	\item SPT-3G and BICEP3 data used there is produced natively on isolatitude (specifically Healpix~\cite{Gorski:2004by}) rings, not on random locations. Yet, this new method, which treats the pixels as a randomly located inside the patch, is also faster than the isolatitude transform truncated to the relevant set of Healpix rings (by a factor of typically 1.5 and 3 respectively), owing to polar optimization and the fine-tuned number of $\Theta$ points\footnote{The small one-off cost of assigning the new spherical coordinates to pixels is not considered in our timings}.
	\item The high-resolution EDFS data is provided natively on a `flat-sky' pixelization, directly amenable to 2D fast Fourier transform. In the cases quoted, we found that our method has comparable or sometimes lower cost than the 2D FFT at the same resolution. 
	\item This method makes possible very efficient calculations on tiny, sub-degree patches as well -- we have used it for efficient theoretical cluster-scale lensing mass reconstruction bias calculations~\cite{Saha:2023bva} in spherical geometry.
\end{itemize}

\section{Conclusions}\label{sec:conclusions}
We have presented a novel algorithm for performing partial-sky spherical harmonic transforms, that builds upon the efficient Fourier‑sphere framework of \citet{Reinecke:2023gtp}. By exploiting the locality of small sky patches, we introduced a scheme that reduces the effective band‑limit of the maps dealt with, thereby decreasing the computational cost. The implementation requires only minor modifications to the existing currently best-performing SHT codebase, preserving all the low‑level optimisations of the original library. A sizeable part of the gain comes from polar optimization -- using spherical coordinates such that the centre of the patch coincides with the pole, making the length of all relevant isolatitude rings as short as possible. The other part from using a fine-tuned, effectively minimal, number of rings to compute and interpolate from. Error control is achieved with the help of carefully designed apodization windows (based on Slepian/Kaiser‑Bessel functions), maximally compact in harmonic space, guaranteeing user‑specified tolerances.

Our main motivation was the problem of high-precision CMB lensing reconstruction~\cite{Hirata:2003ka}, mandatory notably for best constraints on a background of primordial gravitational waves from CMB B-modes~\cite{Belkner:2023duz}. In this case, after porting this method to our CMB data analysis toolset (in particular, Wiener-filtering, quadratic and iterated estimators, spectra estimator), we made the entire process very efficient for massively less or in fact often no need at all for high-performance computing resources.

Beyond our CMB-specific applications, the algorithm is ideally suited to any analysis or calculation that demands repeated spherical harmonic transforms on limited sky regions. Because the method removes the necessity for an isolatitude pixelisation (or binning into pixels at all) and for flat‑sky approximations, it can help in a decisive manner ambitious high‑precision work, such as likelihood-based analyses from catalogs~\cite{Wolz:2024dro, BaleatoLizancos:2023jbr} or other data.

The software is available through \texttt{lenspyx}\footnote{https://github.com/carronj/lenspyx}.\\
\acknowledgements
JC is grateful to Antony Lewis for many useful exchanges on the topic. This work was partly supported by STFC grant UKRI1164 and NSF sub-award AWD104277 (SUB00001436).

\appendix
\section{Spherical harmonics and Fourier series}
One possible definition of spherical harmonics is 
\begin{equation}\label{eq:Ylm}
{}_{s}	Y_{\ell m}(\theta, \phi) =\sqrt{ \frac{2\ell + 1}{4\pi}} d^\ell_{m-s}(\theta)e^{im \phi}.
\end{equation}
The Wigner small-$d$ matrix represents in general a rotation by the angle $\theta$ around the $y$-axis, and there is no explicit restriction to this rotation angle. We can thus meaningfully consider Eq.~\ref{eq:Ylm} for arbitrary values of $\theta$. The symmetries of the Wigner rotation small $d$-symbol $$d^{\ell }_{m-s}(\pi+\beta) = (-1)^{m+s}d^{\ell }_{m-s}(\pi-\beta)$$ imply then
\begin{equation}
		{}_{s}Y_{\ell m}(\pi + \beta, \phi) = (-1)^s\: {}_{s}Y_{\ell m}(\pi-\beta, \phi+\pi).
\end{equation}
As discussed in the main text, the sign $(-1)^s$ must be present in order to capture the oscillatory behavior of spin-weighted fields close to the pole.

The Fourier series of the spherical harmonics follows from that of the Wigner d-matrix. The latter may be obtained by replacing the $y$-axis with the $z$-axis (where the rotation is represented by a phase) with the help of a pair of additional rotations~\cite{Risbo1996FourierTransform, Huffenberger:2010hh, Basak:2008pq}:
\begin{equation}
	d^\ell_{ss'}(\theta)= \sum_{m=-\ell}^\ell i^s d^\ell_{ms}(\pi/2)  e^{-im\theta}  d^\ell_{ms'}(\pi/2)(-i)^{s'}.
\end{equation}
Clearly, the band-limit of the Fourier series for the Wigner small-d matrix elements of multipole $\ell$ is $\ell$. Hence, that of the Fourier series of the spherical map is $\lmax$.

Equations~\eqref{eq:cm} and~\eqref{eq:cms} relating spherical to one-dimensional power along great circles directly follow from this relation. Their high-moment approximations from the correspondence of the Wigner symbols to the Chebyshev polynomials and weights of the first and second kind~\cite{Carron:2024mki},
\begin{equation}
	\begin{split}
	\frac{d^\ell_{sm}(\pi/2)}{d^\ell_{0m}(\pi/2)} &\rightarrow T_s(x)= \cos(s \varphi)\\
	\frac{d^\ell_{sm}(\pi/2)}{d^\ell_{1m}(\pi/2)}&\rightarrow U_{s-1}(x)=\frac{\sin(s \varphi)}{\sin \varphi}\\
	\left|d^\ell_{0m}(\pi/2)\right|^2 &\rightarrow \frac{1}{\pi } \frac{2}{\ell} \frac1{\sqrt{1-x^2}} \\
	\left|d^\ell_{1m}(\pi/2)\right|^2 &\rightarrow \frac{1}{\pi } \frac{2}{\ell} \sqrt{1-x^2}. \\
	\end{split}
\end{equation}
The first line is for $\ell + m$ an even number, the second for $\ell + m$ an odd number. Here, $x=m/\ell = \cos\varphi$.

\end{document}